\documentclass[prd,showpacs,amsmath,amssymb,twocolumn]{revtex4}
\usepackage{graphicx,color}

\usepackage{epsfig}
\usepackage{dcolumn}% Align table columns on decimal point
\usepackage{bm}% bold math

\begin{document}

\title{Understanding the newly observed $Y(4008)$ by Belle}

\author{Xiang Liu}\email{xiangliu@pku.edu.cn}
\vspace*{1.0cm}

\affiliation{School of Physics, Peking University, Beijing 100871,
China}

%\affiliation{$^2$ Department of Physics, Nankai University,
%Tianjin 300071, China}

\date{\today}% It is always \today, today,

\begin{abstract}
Very recently a new enhancement around 4.05 GeV was observed by
Belle experiment. In this short note, we discuss some possible
assignments for this enhancement, i.e. $\psi(3S)$ and $D^*\bar{D}^*$
molecular state. In these two assignments, $Y(4008)$ can decay into
$J/\psi\pi^0\pi^0$ with comparable branching ratio with that of
$Y(4008)\to J/\psi\pi^+\pi^-$. Thus one suggests high energy
experimentalists to look for $Y(4008)$ in $J/\psi\pi^0\pi^0$
channel. Furthermore one proposes further experiments to search
missing channels $D\bar{D}$, $D\bar{D}^*+h.c.$ and especially
$\chi_{cJ}\pi^+\pi^-\pi^0$ and $\eta_c\pi^+\pi^-\pi^0$, which will
be helpful to distinguish $\psi(3S)$ and $D^*\bar{D}^*$ molecular
state assignments for this new enhancement.
\end{abstract}


\maketitle

Very recently Belle Collaboration observed an enhancement with
mass $m=4008 \pm 40^{+114}_{-28}$ MeV and width $\Gamma=226 \pm 44
\pm 87$ MeV besides confirming $Y(4260)$ by studying initial state
radiation (ISR) process $e^+e^-\to \gamma_{ISR} J/\psi \pi^+\pi^-$
\cite{Belle-4008}. Belle experiment also indicated that a fit
using two interfering Breit-Wigner shapes describes the data
better than one that uses only the $Y(4260)$ \cite{Belle-4008}. In
this work, we named this new structure as $Y(4008)$.

Recently a series of observations of charmonium like states $X$,
$Y$, $Z$
\cite{3872-first,3872-CDF,3872-D0,3872-Babar,3872-gamma,3872-gamma-Babar,3872-Belle-3875,3872-Babar-3875,3872-rho-CDF,
3872-angular,3872-angular-CDF,Y4260-Babar,Y4260-CLEO,Y4260-CLEO-2,Y4260-Belle,3872-4260,Y4320,X3940,Y3940,Z3930}
is challenging our understanding for non-perturbative QCD. At
present how to understand this new structure is one of intriguing
and challengeable topics.

In this short note, we are dedicated to the discussion of the
possible interpretations for $Y(4008)$.

\section{A possible candidate for $\psi(3S)$?}

In the known charmonium states listed in Particle Data Book, only
the mass of $\psi(4040)$ is close to that of $Y(4008)$ \cite{PDG}.
At present $\psi(4040)$ is usually considered as the candidate for
$\psi(3S)$. The central value of  width of $Y(4008)$ is larger
than that of $\psi(4040)$ around 100 MeV. However, due to the
large error given by Belle experiment, the mass and width of this
new enhancement are consistent with that of $\psi(4040)$.

For $Y(4008)$, Belle experiment also gave
$B(J/\psi\pi^+\pi^-)\cdot \Gamma_{e^+e^-}=5.0\pm
1.4^{+6.1}_{-0.9}$ eV and $12.4 \pm 2.4^{+14.8}_{-1.1}$ eV
corresponding to two solutions in fitting the data
\cite{Belle-4008}. As the candidate of $\psi(3S)$, the decay width
of $\psi(4040)\to e^+e^-$ is $0.86\pm 0.07$ keV \cite{PDG}. Using
the above values, we can roughly estimate $B[Y(4008)\to
J/\psi\pi^+\pi^-]=5.8\times 10^{-3}$ and $1.4\times 10^{-2}$ for
the above two solutions if $Y(4008)$ is  $\psi(3S)$ state. Due to
the large experimental error, the central value of the former one
is not contradict the upper limit of the branching ratio of
$\psi(4040)\to J/\psi \pi^+\pi^-$ ($B[\psi(4040)\to J/\psi
\pi^+\pi^-]<4\times 10^{-3}$) though the former one is slightly
larger than the upper limit of the branching ratio of
$\psi(4040)\to J/\psi \pi^+\pi^-$.

At present, only $Y(4008)\to J/\psi\pi^+\pi^-$ are reported by
Belle \cite{Belle-4008}. If $Y(4008)$ is $\psi(3S)$, $B[Y(4008)\to
J/\psi\pi^0\pi^0]$ is comparable with $B[Y(4008)\to
J/\psi\pi^+\pi^-]$. Thus $Y(4008)$ can be found in
$J/\psi\pi^0\pi^0$ channel.

\begin{figure}[h]
\begin{center}
\scalebox{0.5}{\includegraphics{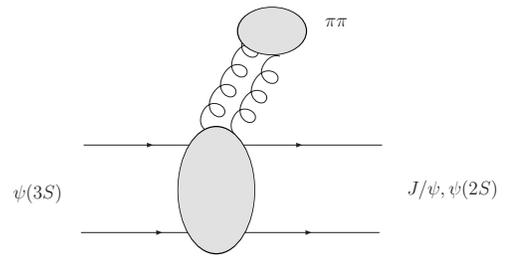}}
\end{center}
\caption{The transition of $\psi(3S)$ to lower states $\psi(nS)$
$(n<3)$ with two pions being emitted.\label{me}}
\end{figure}

Although at present the experiments did not give the measurement
for $\psi(3S)\to J/\psi\pi\pi,\psi(2S)\pi\pi$, the transition of
$\psi(3S)$ to lower states $\psi(nS)$ $(n<3)$ with two pions being
emitted can be solved by the QCD multipole expansion (QCDME)
method proposed by Gottfried, Yan and Kuang \cite{qcdme}, which is
depicted by Fig. \ref{me}.
 In a recent work
\cite{ke}, Ke et al. calculated the transitions of $\psi(3S)\to
\psi(nS)\pi\pi$, and obtain
\begin{eqnarray}
&&\Gamma[\psi(3S)\to J/\psi\pi\pi]=589.91\;\rm{keV},\\
&&\Gamma[\psi(3S)\to \psi(2S)\pi\pi]=14.96\;\rm{keV},
%&&B[\psi(3S)\to J/\psi\pi^+\pi^-]=**,\\
%&&B[\psi(3S)\to J/\psi\pi^0\pi^0]=**,\\
%&&B[\psi(3S)\to \psi(2S)\pi^+\pi^-]=**,\\
%&&B[\psi(3S)\to \psi(2S)\pi^0\pi^0]=**,
\end{eqnarray}
by adopting the Cornell potential $V(r)=-\frac{\kappa}{r}+br$
\cite{cornel potential} and
\begin{eqnarray}
&&\Gamma[\psi(3S)\to J/\psi\pi\pi]=12.38\;\rm{keV},\\
&&\Gamma[\psi(3S)\to \psi(2S)\pi\pi]=8.84\;\rm{keV},
%&&B[\psi(3S)\to J/\psi\pi^+\pi^-]=**,\\
%&&B[\psi(3S)\to J/\psi\pi^0\pi^0]=**,\\
%&&B[\psi(3S)\to \psi(2S)\pi^+\pi^-]=**,\\
%&&B[\psi(3S)\to \psi(2S)\pi^0\pi^0]=**,
\end{eqnarray}
by adopting a modified Cornell potential which includes a
spin-related term \cite{cornel potential-m}
$$V(r)=-\frac{\kappa}{r}+br+\frac{8\pi\kappa}{3m^2_{q}}\delta_{\sigma}(r)\mathbf{S}_q\cdot \mathbf{S}_{\bar q}+V_0,$$
where $\delta_{\sigma}(r)=(\frac{\sigma}{\sqrt{\pi}})^3
e^{-\sigma^2 r^2}$ and $V_0$ is the zero-point energy (for more
detail, see Ref. \cite{ke}). The above numerical results by two
potential show that there exists large uncertainty for the
estimate of $\psi(3S)\to J/\psi\pi\pi$ by QCDME method, which is
indicated in Ref. \cite{ke}. However the estimate of $\psi(3S)\to
\psi(2S)\pi\pi$ without spin-related term is consistent with that
with spin-related term. If we trust the estimate of $\psi(3S)\to
\psi(2S)\pi\pi$ by QCDME method, it is hopeful to search $Y(4008)$
in $\psi(2S)\pi\pi$ channel in future experiments.

%Although the results obtained by two potential are different, the
%order of magnitude of $\psi(3S)\to J/\psi\pi\pi$ is about
%$10^{-5}\sim 10^{-3}$. The value of $B[Y(4008)\to
%J/\psi\pi^+\pi^-]=5.8\times 10^{-3}$ is still consistent with the
%upper limit of that obtained by QCDME.

Furthermore, if $Y(4008)$ is $\psi(3S)$, we know that
$J/\psi\pi^+\pi^-$ is not its main decay channel. $Y(4008)$ can
mainly decay into $D\bar{D}$ and $D\bar{D}^*+h.c.$. Due to the
fact that $Y(4008)$ is of wide decay width with about 200 MeV,
$Y(4008)$ can also decay into $D^*\bar{D}^*$ through its mass
tail.

%In a summary, for finally determining if $Y(4008)$ is $\psi(3S)$,
%some efforts are still needed: (1) To decrease the error of mass and
%width in further experiments; (2) To Search the $Y(4008)\to
%J/\psi\pi^0\pi^0, \psi(2S)\pi\pi$; (3) To search the missing
%channels $D\bar{D}$, $D\bar{D}^*+h.c.$ and $D^*\bar{D}^*$.

\section{A $D^*\bar{D}^*$ molecular state?}

There has been a long history about the molecular structure of
hadrons. To explain some phenomena which are hard to find natural
interpretations in the canonical framework, people have tried to
search for new structure beyond it. The molecular structure is one
of the possible candidates.

\begin{figure}[h]
\begin{center}
\scalebox{0.7}{\includegraphics{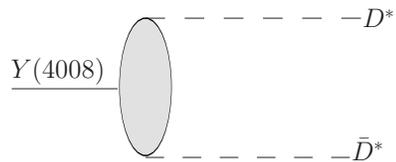}}
\end{center}
\caption{The diagrams depicting the $Y(4008)\to D^*\bar{D}^*$ decay.
\label{fallapart}}
\end{figure}

Because the mass of $Y(4008)$ is close to the threshold of
$D^*\bar{D}^*$, and $Y(4008)$ is of about 200 MeV wide width, thus
$Y(4008)$ can be assumed as a $D^*\bar{D}^*$ molecular state. In
the history, Okun and Voloshin studied the interaction between
charmed mesons and proposed possibilities of the molecular states
involving charmed quarks \cite{Okun}. Rujula, Geogi and Glashow
once suggested $\psi(4040)$ as a $D^*\bar{D}^*$ molecular state
\cite{RGG}. In Ref. \cite{voloshin-1,voloshin}, Dubynskiy and
Voloshin proposed that there exists a possible new resonance at
the $D^*\bar{D}^*$ threshold. Because $Y(4008)$ is observed along
with $Y(4260)$ which is of $J^{PC}=1^{--}$, thus the most possible
quantum number of $Y(4008)$ is $J^{PC}=1^{--}$. Furthermore
$Y(4008)$ must be a p-wave $D^*\bar{D}^*$. At present one can not
use the experimental information to determine the quantum number
$I^{G}$ of $Y(4008)$. Thus $Y(4008)$ can be isosinglet state with
$I^{G}=0^-$ or isovector state with $I^{G}=1^+$. If $Y(4008)$ is a
$D^*\bar{D}^*$ molecular state, $Y(4008)$ falls apart into
$D^*\bar{D}^*$ by its mass tail, which is depicted in Fig.
\ref{fallapart}.
In the following we will discus its other possible decay modes.

\vspace{0.5cm}

(i) $Y(4008)$ as an isoscalar $D^*\bar{D}^*$ molecular state

\vspace{0.3cm}

\begin{figure}[h]
\begin{center}
\scalebox{0.7}{\includegraphics{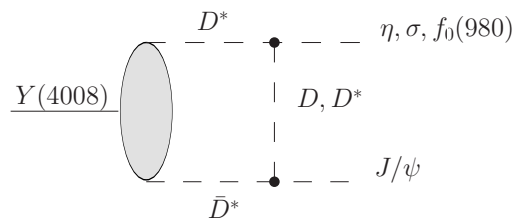}}
\end{center}
\caption{The diagrams depicting the $Y(4008)\to
J/\psi\eta,J/\psi\sigma,J/\psi f_0(980)$ decays. \label{eta}}
\end{figure}

\begin{figure}[h]
\begin{center}
\scalebox{0.7}{\includegraphics{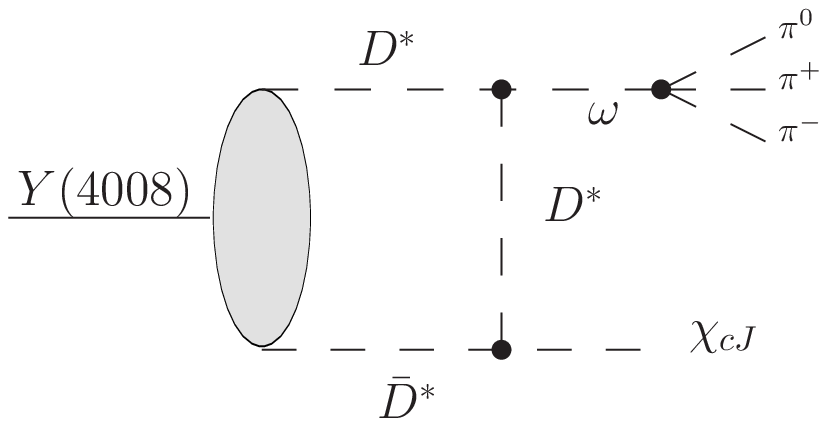}}\\\hspace{0.2cm}\\
\scalebox{0.7}{\includegraphics{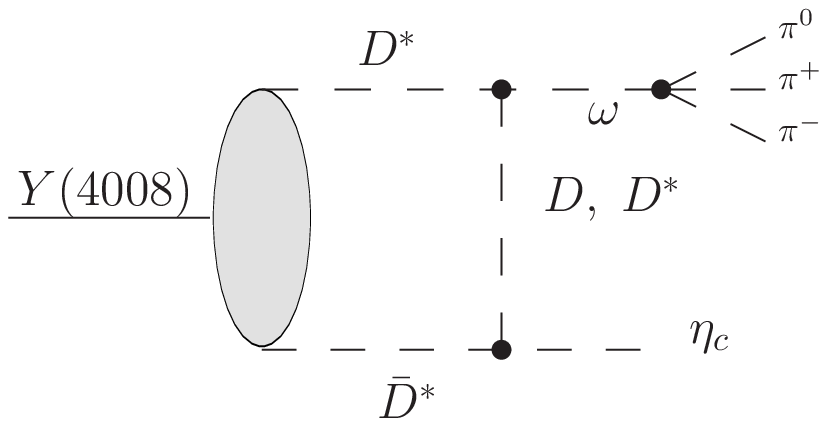}}\\
\end{center}
\caption{The diagrams depicting the $Y(4008)\to
\chi_{cJ}\omega,\eta_c\omega$ decays. \label{omega}}
\end{figure}

By the $D^*\bar{D}^*$ recattering effect, $Y(4008)$ with $I^G=0^-$
can decay into $J/\psi+\eta$, $J/\psi+\sigma$ and
$J/\psi+f_0(980)$ by the mechanism depicted in Fig. \ref{eta}, and
into $\chi_{cJ}\omega\; (J=0,1,2),\;\eta_c\omega$ by Fig.
\ref{omega}. Here $J/\psi$ can be also replaced as $\psi(2S)$ and
$\psi(3770)$.
By the same mechanism, $Y(4008)$ also decay into $D\bar D$ and
$D\bar{D^*}+h.c.$ by exchanging $\pi$ and $\rho$ mesons between
$D^*$ and $\bar{D}^*$. In fact, as secondary decay, the branching
ratio of $Y(4008)\to D^*\bar D^*\to D\bar{D}, D\bar{D}^*+h.c.$ is
comparable with that of $Y(4008)\to D^*\bar D^*\to J/\psi\eta,
J/\psi\omega$.

Because $\sigma$ and $f_{0}(980)$ dominantly decay into $\pi\pi$,
thus, according to the isospin symmetry, one can roughly estimate
\begin{eqnarray}
\frac{B[Y(4008)\to J/\psi\pi^0\pi^0]}{B[Y(4008)\to
J/\psi\pi^+\pi^-]}\sim\frac{1}{2}.
\end{eqnarray}
Furthermore the decay mechanism depicted by Fig. \ref{eta} can be
test in further experiments by analyzing the $\pi\pi$ invariant mass
spectrum. If this mechanism is correct, the $\pi\pi$ invariant mass
distribution should show the signature of $\sigma$ or $f_0(980)$.

The branching ratio of $\omega\to \pi^+\pi^-\pi^0$ is almost
89.1\%, thus $\omega$ to $\pi^+\pi^-\pi^0$ is overwhelming.
$\chi_{cJ}\pi^+\pi^-\pi^0$ and $\eta_c\pi^+\pi^-\pi^0$ are
expected as special and main decay modes of $Y(4008)$. Meanwhile
$\omega$ also decay into $\pi^+\pi^-$ and $\pi^0\gamma$ with the
branching ratio $B(\omega\to \pi^+\pi^-)=1.7\%$ and $B(\omega\to
\pi^0\gamma)=8.9\%$ respectively \cite{PDG}. Thus
$\chi_{cJ}\pi^+\pi^-$, $\chi_{cJ}\pi^0\gamma$,
$\eta_{c}\pi^+\pi^-$, $\eta_{c}\pi^0\gamma$ are important decay
modes for $Y(4008)$.

The typical decay modes of $Y(4008)$ with the assignment of
$D^*\bar{D}^*$ molecular state ($I^G (J^{PC})=0^-(1^{--})$) mainly
include $J/\psi\eta$, $J/\psi \pi\pi$, $\chi_{cJ}\pi^+\pi^-\pi^0$,
$\chi_{cJ}\pi^0\gamma$, $\chi_{cJ}\pi^+\pi^-$,
$\eta_{c}\pi^+\pi^-\pi^0$, $\eta_{c}\pi^0\gamma$,
$\eta_{c}\pi^+\pi^-$, $D\bar{D}$, $D\bar{D}^*+h.c.$. As one of the
main decay modes, $\chi_{cJ}(\eta_c)\pi^{+}\pi^-\pi^0$ should be
seen if $Y(4008)$ is a $D^*\bar{D}^*$ molecular state with
$I^G=0^-$. However there exists the difficulty to distinguish
$Y(4008)\to\chi_{cJ}(\eta_c)\pi^0\gamma$ in the experiment. Here
index $J$ of $\chi_{cJ}$ can be $0,1,2$.

\vspace{0.5cm}

(ii) $Y(4008)$ as an isovector $D^*\bar{D}^*$ molecular state

\begin{figure}[htb]
\begin{center}
\scalebox{0.6}{\includegraphics{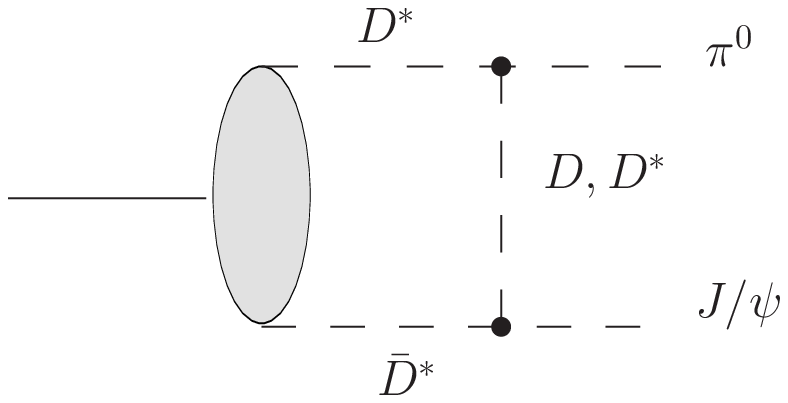}}\\\hspace{0.2cm}\\
\scalebox{0.6}{\includegraphics{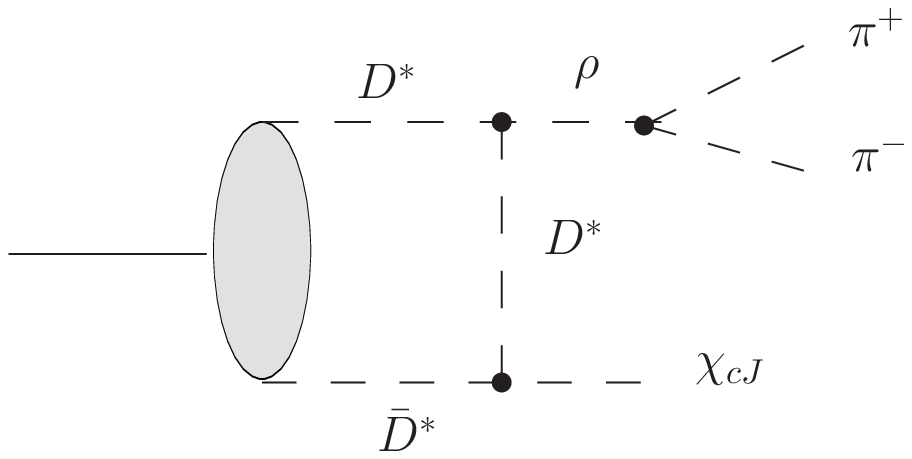}}\\\hspace{0.2cm}\\
\scalebox{0.6}{\includegraphics{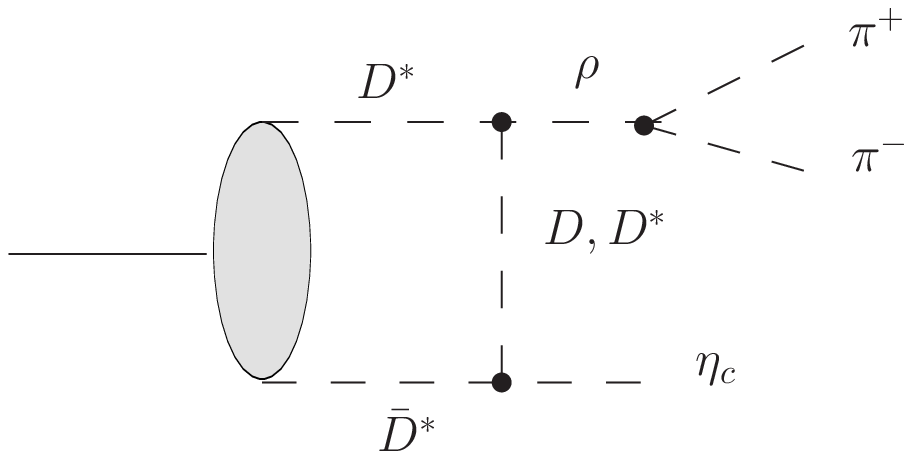}}\\
\end{center}
\caption{The diagrams depicting the $Y(4008)\to
\chi_{cJ}\rho,\eta_c\rho$ decays. \label{rho}}
\end{figure}

For isovector $D^*\bar{D}^*$ molecular state, $Y(4008)$ can decay
into $\pi^0 J/\psi$, $\rho^0\chi_{cJ}\;(J=0,1,2)$ and
$\rho^0\eta_{c}$, which are depicted in Fig. \ref{rho}.
The branching ratio of $\rho^0\to \pi^+\pi^-$ is almost $100\%$
\cite{PDG}. Thus the typical decay modes of $Y(4008)$ as an
isovector $D^*\bar{D}^*$ molecular state are $\pi^0 J/\psi$,
$\chi_{cJ}\pi^+\pi^-$ and $\eta_{c}\pi^+\pi^-$. Besides these
decays, of course $Y(4008)$ can also decay into $D\bar{D}$ and
$D\bar{D}^*+h.c.$. Because $J/\psi\pi^+\pi^-$ is forbidden for an
isovector $D^*\bar{D}^*$ molecular state, thus one can exclude the
assignment of isovector $D^*\bar{D}^*$ molecular state for
$Y(4008)$.

\vfill
\section{Brief Conculsion}

In the above sections, we discuss possible assignments for
$Y(4008)$: $\psi(3S)$ and $D^*\bar{D}^*$ molecular state. In these
two possible pictures, one finds that the branching ratio of
$Y(4008)\to J/\psi\pi^0\pi^0$ is comparable with that of
$Y(4008)\to J/\psi\pi^+\pi^-$. Thus one suggests further
experiments to search $Y(4008)$ in $J/\psi\pi^0\pi^0$ invariant
mass distribution.

How to distinguish these two assignments becomes a key problem. In
the following we will illustrate the differences of $Y(4008)$ decays
for two assignments, which will be helpful to distinguish $\psi(3S)$
and $D^*\bar{D}^*$ molecular state pictures:

(1) Search for $D\bar{D}$, $D\bar{D}^*+h.c.$ decay channels. If
$Y(4008)$ is $\psi(3S)$, $D\bar{D}$, $D\bar{D}^*+h.c.$ are main
decay channels. If $Y(4008)$ is a $D^*\bar{D}^*$ molecular state,
$D\bar{D}$, $D\bar{D}^*+h.c.$, as the secondary decay modes, are
comparable with $\chi_{cJ}\pi^+\pi^-\pi^0$ and $\eta_c
\pi^+\pi^-\pi^0$. Thus one suggests experiments to search these
missing decay channels.

(2) Search for $\chi_c\pi^+\pi^-\pi^0$ and $\eta_c
\pi^+\pi^-\pi^0$ decay channels. In the picture of $D^*\bar{D}^*$
molecular state, $\chi_{cJ}\pi^+\pi^-\pi^0$ and $\eta_c
\pi^+\pi-\pi^0$ are main decay modes. However, as $\psi(3S)$,
besides decaying to $D\bar{D}$, $D\bar{D}^{*}+h.c.$ and
$D^*\bar{D}^*$, $Y(4008)$ mainly decays into $J/\psi \pi\pi$. It
will be a decisive factor to distinguish $\psi(3S)$ and
$D^*\bar{D}^*$ molecular state assignments if the $Y(4008)\to
\chi_{cJ} \pi^+\pi^-\pi^0,\eta_{c} \pi^+\pi^-\pi^0$ can be found
in further experiments. One strongly urges our experimental
colleague to design more accurate experiments to find $Y(4008)\to
\chi_{cJ} \pi^+\pi^-\pi^0,\eta_{c} \pi^+\pi^-\pi^0$.

\vfill

%%%%%%%%%%%%%%%%%%%%%%%%%%%%%%%%
\section*{Acknowledgments}
%%%%%%%%%%%%%%%%%%%%%%%%%%%%%%%%

We thank H.W. Ke for useful communication about their work. We
also thank Prof. S.L. Zhu for interesting discussions and useful
suggestions. This project was supported by the National Natural
Science Foundation of China under Grants 10421503, 10625521
10705001, and the China Postdoctoral Science foundation under
Grant No 20060400376.


\begin{thebibliography}{99}
\bibitem{Belle-4008}Belle Collaboration, C.Z. Yuan et al.,
arXiv: 07072541v1 [hep-ex].

%%%%%   exp
\bibitem{3872-first}Belle Collaboration, S.K. Coi et al., Phys. Rev. Lett. {\bf 91}, 262001 (2003).
\bibitem{3872-CDF}CDF Collaboration, D. Acosta et al., Phys. Rev. Lett. {\bf 93}, 072001 (2004).
\bibitem{3872-D0}D0 Collaboration, V.M. Abazov et al., Phys. Rev. Lett. {\bf 93}, 162002 (2003).
\bibitem{3872-Babar}Babar Collaboration, B. Aubert et al., Phys. Rev. {\bf D 71}, 071103 (2005);
%B. Aubert et al., Phys. Rev. {\bf D 73}, 011103 (2006).
\bibitem{3872-gamma}Belle Collaboration, K. Abe et al., arXiv: hep-ex/0505037.
\bibitem{3872-gamma-Babar}Babar Collaboration, B. Aubert, Phys. Rev. {\bf D 74}, 071101
(2006).
\bibitem{3872-Belle-3875}Belle Collaboration, G. Gokhroo et al., Phys. Rev. Lett. {\bf 97}, 162002
(2006).
\bibitem{3872-Babar-3875}Babar Collaboration, talk given by P.
Grenier in Moriond QCD 2007, 17-24 March,
http://moriond.in2p3.fr/QCD/2007/SundayAfternoon/\\Grenier.pdf.

\bibitem{3872-rho-CDF}CDF Collaboration, A. Abulencia et al., Phys. Rev. Lett. {\bf 96}, 102002 (2006).
\bibitem{3872-angular}Belle Collaboration, K. Abe et al., arXiv: hep-ex/0505038.
\bibitem{3872-angular-CDF} CDF Collaboration, A. Abulencia, Phys. Rev. Lett {\bf 98}, 132002
(2007).

%%%%%%%%%%%%%%%%%%%%  Y(4260)   %%%%%%%%%%%%%%

\bibitem{Y4260-Babar}Babar Collaboration, B. Aubert et al.,  Phys. Rev. Lett. {\bf 95}, 142001 (2005).
\bibitem{Y4260-CLEO} CLEO Collaboration, T. E. Coan et al.,Phys. Rev. Lett. {\bf 96},
162003 (2006).
\bibitem{Y4260-CLEO-2}CLEO Collaboration, Q. He et al., Phys. Rev. {\bf D 74}, 091104 (2006).
\bibitem{Y4260-Belle}Belle Collaboration, K. Abe et al., arXiv:
hep-ex/0612006.


\bibitem{3872-4260}Babar Collaboration, B. Aubert et al.,  Phys. Rev. {\bf D 73}, 011101 (2006).

%%%%%%%%%%%%%%%%%%    Y4320   %%%%%%%%%%%%%%%%

\bibitem{Y4320}Babar Collaboration, B. Aubert, et al., Phys. Rev. Lett. {\bf 98}, 212001
(2007).
%%%%%%%%%%%%%%%%%%%%%   Z(3930)  %%%%%%%%%%%%%%%%%%%%%%%%%%%%%%
\bibitem{X3940}Belle Collaboration, K. Abe et al., arXiv: hep-ex/0507019.

\bibitem{Y3940}Belle Collaboration, S.K. Choi et al., Phys. Rev. Lett. {\bf94}, 182002 (2005).
\bibitem{Z3930}Belle Collaboration, S. Uehara et al., Phys. Rev. Lett. {\bf 96}, 082003 (2006).



\bibitem{PDG}W.M. Yao et al., Particle Data Group, J. Phys. G {\bf 33}, 1
(2006).

\bibitem{qcdme}K. Gottfried, Phys. Rev. Lett. {\bf 40}, 598(1978);
Y.P. Kuang and T.M. Yan, Phys. Rev. {\bf D 24}, 2874(1981); Y.P.
Kuang , Front. Phys. China {\bf 1}, 19(2006); T.M. Yan, Phys. Rev.
{\bf D 22}, 1652(1980); Y.P. Kuang, Y.P. Yi and B. Fu, Phys. Rev. {\bf
D 42}, 2300(1990).

\bibitem{ke}H.W. Ke, J. Tan, X.Q. Hao and X.Q. Li,
Phys. Rev. {\bf D 76}, 074035 (2007).

\bibitem{cornel potential}E. Eichten, K. Gottfried, T. Kinoshita, K.D. Lane and T.M. Yan , Phys. Rev. {\bf D 17},
3090 (1978); ibid, {\bf D 21}, 203(1980).
\bibitem{cornel potential-m}T. Barnes , S. Godfrey and E.S. Swanson, Phys. Rev. {\bf D 72}, 054026 (2005).

\bibitem{Okun} M.B. Voloshin and L.B. Okun, JETP Lett. {\bf 23}, 333 (1976).

\bibitem{RGG}A.D. Rujula, H. Georgi and S.L. Glashow, Phys. Rev.
Lett. {\bf 38}, 317 (1977).

\bibitem{voloshin-1}M.B. Voloshin, arXiv:hep-ph/0602233.
\bibitem{voloshin}S. Dubynskiy and M.B. Voloshin, Mod. Phys. Lett. {\bf A 21}, 2779
(2006).




\end{thebibliography}
\end{document}